\documentclass[aip,cha,amsmath,amssymb,reprint,floatfix]{revtex4-1}

\usepackage{graphicx}
\usepackage{dcolumn}
\usepackage{bm}
\usepackage[utf8]{inputenc}
\usepackage[T1]{fontenc}
\usepackage{mathptmx}
\usepackage{etoolbox}  
\usepackage{xcolor}    
\usepackage[normalem]{ulem}
\usepackage{soul}

\makeatletter
\def\@email#1#2{%
 \endgroup
 \patchcmd{\titleblock@produce}
  {\frontmatter@RRAPformat}
  {\frontmatter@RRAPformat{\produce@RRAP{*#1\href{mailto:#2}{#2}}}\frontmatter@RRAPformat}
  {}{}
}%
\makeatother
\begin{document}

\preprint{AIP/123-QED}
\title[]{Percolation with coupled lasers:\\effect of non-linearities on the phase transition}
\author{Simon Mahler}
\email{smahler@stevens.edu}
\thanks{These authors contributed equally to this work.}
\affiliation{Department of Physics of Complex Systems, Weizmann Institute of Science, Rehovot 761001, Israel.}
\affiliation{Department of Biomedical Engineering, Stevens Institute of Technology, Hoboken, NJ 07030, USA.}
\author{Nikita Stroev}
\thanks{These authors contributed equally to this work.}
\affiliation{Department of Physics of Complex Systems, Weizmann Institute of Science, Rehovot 761001, Israel.}
\author{Mahmoud Abu Rmilah}%
\thanks{These authors contributed equally to this work.}
\affiliation{Department of Physics of Complex Systems, Weizmann Institute of Science, Rehovot 761001, Israel.}
\author{Asher Friesem}%
\affiliation{Department of Physics of Complex Systems, Weizmann Institute of Science, Rehovot 761001, Israel.}
\author{Nir Davidson}%
\affiliation{Department of Physics of Complex Systems, Weizmann Institute of Science, Rehovot 761001, Israel.}

\date{\today}

\begin{abstract}
Controlled experimental studies of percolation are challenging due to difficulties in tuning site connectivity, isolating local interactions, and mitigating finite-size effects. In this work, we experimentally investigate percolation with a platform of coupled lasers, where connectivity, interaction strength, and system size can be controlled. Using a square array of $100$ lasers with astronomical number of possible cluster configurations, we show that the emergence of a percolating cluster corresponds to the onset of phase locking among the lasers. We also show that the percolation probability undergoes a second-order alike transition as a function of the site-occupation probability, with a threshold consistent with classical theoretical predictions. Surprisingly, we find that at low pump level, amplified mode competition (nonlinear regime) alters the effective behavior of the lasing sites and modify the nature of the percolation transition. The experimental results are interpreted by the means of a theoretical toy model with connectivity rules to the classical percolation. 
\end{abstract}

\maketitle
\section{Introduction}
\label{Introduction}
Percolation theory involves a critical transition of a system between a non-connected to a fully connected state of a system~\cite{StaufferAharony2018,Broadbent57}. In percolation, phase transition refers to the emergence of a system-wide connected cluster at a critical site-occupation probability, where the size of the largest connected cluster grows from microscopic (disconnected clusters) to macroscopic (extensive), with the transition becoming sharper as the system size increases (finite-size effects) \cite{Saberi15,Ziff2010}.

Generally, the system is a network (array) of sites, where sites (nodes) can link (bond) with their neighbors depending on their connections state~\cite{Grimmett1999Percolation}. Percolation theory deals with threshold phenomena, where near the critical occupation probability (for either sites or bonds), the system exhibits critical behavior, i.e., the absence of a characteristic length scale and the emergence of long-range correlations. For example, the order parameter  and correlation length  follow power-law, characterized by critical exponents, which quantify how these observables diverge or vanish near the transition
~\citep{StaufferAharony2018,Grimmett1999Percolation}. These exponents are universal, meaning they depend only on general properties such as the dimensionality of the system, rather than microscopic details~\cite{Hughes1996CriticalExponents,Saberi15}. 

Experimental realizations of percolation phenomena span a diverse range of classical and  quantum physical systems. Comprehensive reviews~\citep{sarikhani2022unified,Saberi15} have surveyed percolation in condensed matter, biological systems, and network systems. Notably, percolation has been explored in networks of one-dimensional objects, where both Monte Carlo simulations and experimental observations reveal critical connectivity thresholds and transport behavior~\citep{langley2018percolation}. Directed percolation with universal scaling near the transition~\citep{takeuchi2009experimental} was experimentally demonstrated in turbulent liquid crystals, while polymer networks have exhibited classical percolation transitions tied to conductivity~\citep{andrade1996percolation}. The scaling refers to power-law dependencies of measurable quantities such as cluster size, correlation length, or order parameter. Additional studies include percolation in random networks~\citep{feinerman2017experimental} and more recently direct observations of quantum percolation dynamics in photonic chip lattices~\citep{feng2023direct}. Also, numerical simulations are continually refining analytic models of percolation, yet challenges remain~\citep{Araujo14}.

Despite all the advances, experimental platforms enabling fully tunable site percolation with coherent coupling and nonlinear site control have not fully developed, particularly those capable of tracking both intensity-based connectivity and phase coherence of arrays. Moreover, the challenges are compounded in two dimensions, where connected clusters of occupied sites (lattice animals) grow so rapidly in number with increasing size that they overwhelm the combinatorial landscape~\citep{Iwan2000,Redelmeier1981}, making it difficult to probe percolation mechanisms in a controlled setting. Recent efforts to count lattice animals, \citep{Redelmeier1981,Polyominoes59, Barequet2024}, are currently limited to $12$ connected sites \citep{Polyominoes59, Barequet2024}.

In this paper, we introduce a photonic percolation simulator based on an array of coupled lasers with a tunable and programmable site configuration. Specifically, we demonstrate percolation within a square array of $100$ coupled lasers, exploring a wide variety of lattice-animal configurations. Using a digital spatial light modulator (SLM) embedded in a degenerate cavity laser (DCL)~\cite{Nixon13}, we precisely control the location, orientation, amplitude, and phase of individual lasing sites, like programmable disorder ~\cite{pando2024synchronization,Mahler22}. By tuning the lasing pump strength $R$, we obtain an additional control parameter for probing nonlinear lasing effects on the percolation transition and demonstrate nonlinear renormalization of the effective percolation threshold.

Surprisingly, at low pump strength, where lasing nonlinearities are the strongest mode competition is strong (nonlinear regime, see Supplementary Material Fig. S2), we find that the effective percolation threshold is significantly increased, due to selective suppression of small clusters. This is a distinctive observable feature of our nonlinear system.  We also measure increased correlations in site occupations at low pumps (high nonlinearities). Finally, we observe and explain how the coherence of the coupled laser system is related to the nonlinear percolation transition. Our numerical simulations of the dynamical nonlinear laser equations agree well with all these experimental measurements. 

To explain our results we introduce a simple theoretical \emph{survival-game} toy model that maps the pump strength of the sites onto a neighbor-dependent set of connectivity rules (a framework not reported in linear percolation and laser literature). This toy model includes threshold shifts and scaling under nonlinearity, offering a simple yet powerful tool that bridges percolation theory with nonlinear laser dynamics and could be adapted in other network systems.

\section{Experimental Arrangement}
\label{Experimental Setup}

\begin{figure}
\centering
\includegraphics[width=0.45\textwidth]{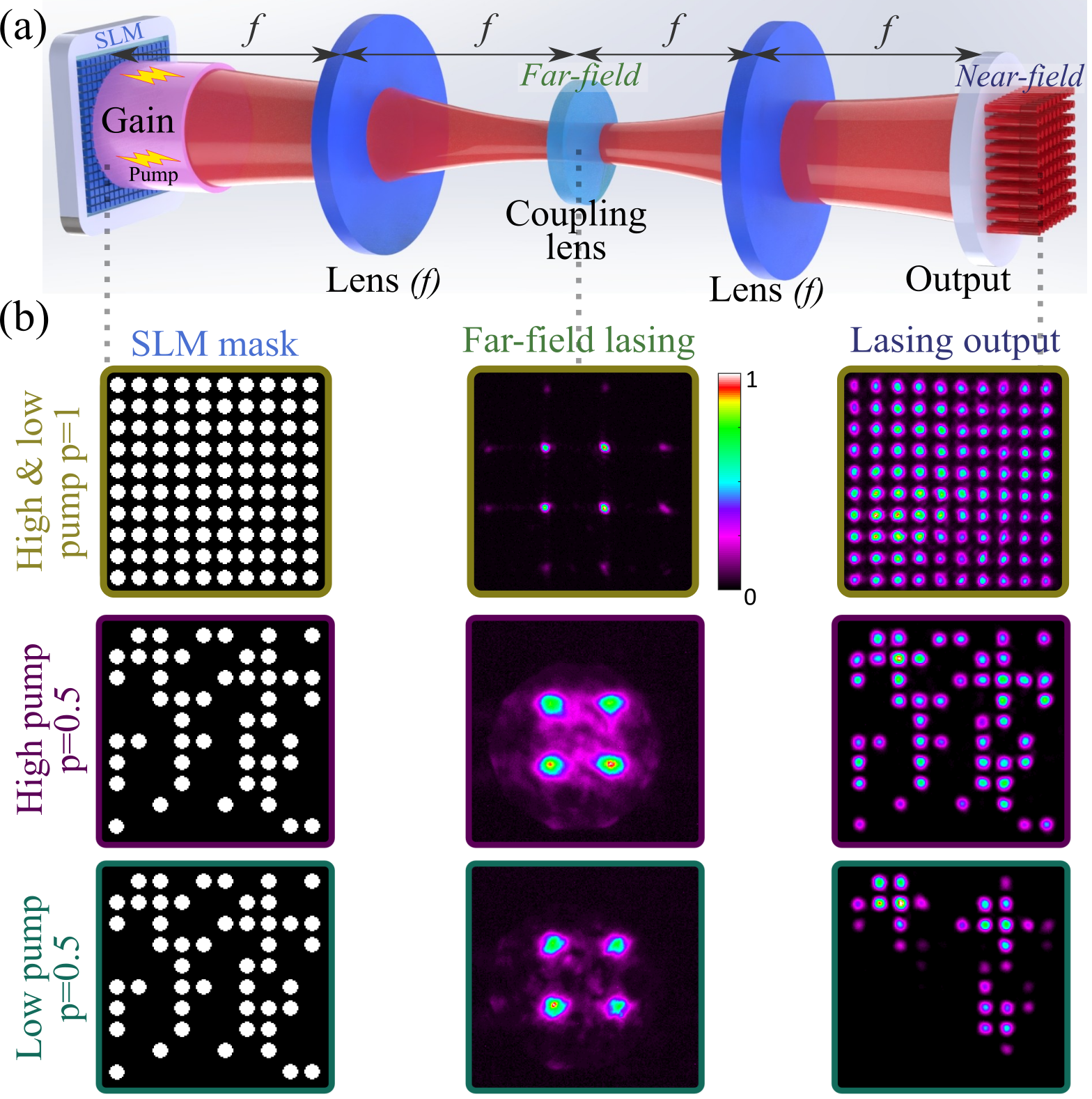}
\caption{Percolation with coupled lasers. (a) Experimental arrangement comprised of a degenerate cavity laser (DCL) with an intra-cavity 4f telescope, a spatial light modulator (SLM) serving as laser array generator, and a far-field coupling lens that couples nearest-neighbors lasers. (b) SLM mask (left), far-field (center) and near-field (right) lasing outputs intensity distributions at high and low pump values for site probability $p=1$ (top) and $p=0.5$ (middle and bottom). At low pump, only large clusters are lasing, and the coherence of the array, manifested by the sharpness of the far-field diffraction peaks, is slightly improved for lower $p$.}
\label{fig:Fig_1_Exp_sketch}
\end{figure}

Our experimental arrangement for implementing square arrays of 100 lasers with independent intensities and phases and a programmable site probability $p$ is shown in Fig.~\ref{fig:Fig_1_Exp_sketch}(a). It is based on a degenerate cavity laser (DCL)~\cite{Arnaud69}, composed of a 4f telescope placed between a $90 \%$ reflectivity  output coupler and a computer-controlled spatial light modulator (SLM) [Hamamatsu LCOS X10468-03]. The SLM serves as an end-mirror with programmable reflectivity that can generate any desired complex multi-mode or structured lasing field \citep{Tradonsky21}. Here, we used the SLM to generate the underlying lattice on which the percolation process is implemented, by initializing each site (i.e, laser) with a probability $p$~\citep{Hammersley1980,Chayes2000,Vizi2018}.

The 4f telescope ensures that the transverse field distribution is self-imaged after each round trip, making all transverse modes degenerate in loss and frequency, and allowing  them to be simultaneously supported~\cite{Nixon13,Mahler21,Mahler22}.  
The two telescope lenses were anti-reflective coated plano-convex lenses of 5.08 cm diameters and f = 25 cm focal length. The gain medium was a solid-state Nd:YAG rod of 0.95 cm diameter and 10.9 cm length lasing at wavelength $\lambda=1064$ nm. The optical pump was provided by two xenon flash lamps pulsing at 1 Hz with $100$ $\mu$s quasi-continuous-wave  duration, to minimize thermal lensing. We varied the optical pump power to be between $R=1.1$ and $R=1.4$ times the lasing threshold. The determination of the pump power at lasing threshold $P_{th}(p)$ and its dependence on the site occupation probability $p$ are detailed in the Supplementary Material. We find that $P_{th}$ decreases with increasing $p$, an effect that was accounted for in the experiments (see Supplementary Material).

The intensity distribution at the end-mirrors (near-field plane) and at their common focal plane (far-field plane)  were recorded simultaneously~\citep{Chriki22} on different regions of a CMOS camera (Ximea MQ013MG-E2). The SLM was programmed to generate a highly reflective mask of square array of circular mirrors with period of $a=300$ $\mu$m. The diameter of each mirror was chosen to be 200 $\mu$m, to support an independent single-mode laser centered on its site. Each  mirror was programmed to have probability $p$ to be highly reflective. The SLM reflectivity mask with $p=1$ and the resulting lasing output intensity distribution, depicted in Fig.~\ref{fig:Fig_1_Exp_sketch}(b,top), revealed a unit filling square array of 100 lasers.   

Strong negative coupling between nearest-neighbor lasers was achieved by placing a spherical lens with focal length $f_{C}=1.5$ m at the far-field plane, which emulates Talbot diffraction at half the Talbot distance $Z_{T}=a^2/(2\lambda)$ ~\cite{Chavel84,Leger84,Mahler19_2}. The resulting far-field diffraction pattern (Fig.~\ref{fig:Fig_1_Exp_sketch}b, top) was composed of 4 sharp diffraction peaks indicating that the laser array was highly coherent and out-of-phase locked ~\cite{Tradonsky17,Naresh22,Mahler20}. Specifically, the sharpness of the far-field diffraction peaks quantified the number of phase-locked lasers~\cite{Mahler20_3} as the square of the ratio between the diffraction peaks spacing  $D$ over their full width at half maximum (FWHM) ~\cite{Mahler20_3}:

\begin{equation}
M_{locked}=(D/FWHM)^2.
\label{eq:M_locked}
\end{equation}

For unit site probability $p=1$ (Fig.~\ref{fig:Fig_1_Exp_sketch}b), the far-field intensities distribution had darkness in the center and the number of phase locked lasers was $M_{locked}>50$ for all values of R indicating that most of our $100$ lasers in the array are out-of-phase locked  ~\cite{Tradonsky17,Naresh22,Mahler20}.

Next, we set the SLM mask for probability $p=0.5$ and explored two different pump regimes: high pump (R=1.4, well above threshold) and low pump (R=1.1, near the lasing threshold). The results in Fig.~\ref{fig:Fig_1_Exp_sketch}b clearly show that for high pump, the near-field lasing intensity distribution closely matches the mask pattern on the SLM (Fig.~\ref{fig:Fig_1_Exp_sketch}b, middle row). This inidcates that all lasers supported by the reflective parts of the SLM mask indeed lase. The corresponding far-field intensity distribution exhibits the characteristic pattern of out-of-phase locked lasers ~\cite{Tradonsky17,Naresh22,Mahler20}. While the diffraction peaks indicate that many lasers in the array are phase-locked, the presence of a diffuse background signal (absent for $p=1$) reveals that some isolated clusters are not phase-locked as they are disconnected from the main clusters~\cite{Mahler20_3}.

For low pump, a different behavior emerges: only larger main clusters survive, while smaller clusters fail to lase ((Fig.~\ref{fig:Fig_1_Exp_sketch}b, bottom row)) . Here, a cluster is defined as a group of nearest-neighbor lasers that are all connected. This behavior arises because mode competition becomes more pronounced near the lasing threshold. Strong nearest-neighbor coupling favors larger interconnected clusters, where lasers can mutually support each other, and in turn, lower their overall loss. Consequently, for low pump conditions, the reduced available gain prevents small clusters from reaching the lasing threshold, and they fail to  lase. Such selection of the largest clusters, becomes more pronounced as the pump approaches the lasing threshold. In this inherently non-linear regime, the interplay of mode competition and nearest-neighbor coupling leads to a non-linear selection of larger lasing clusters, where the output is not directly proportional to the input pump. 

Although the overall number of lasing lasers is lower for low-pump, the sharpness of the far-field diffraction peaks is similar to that of the high-pump case, indicating that the number of phase locked lasers is similar. Also, the diffuse background signal is weaker for low pump, suggesting that fewer lasers in the small clusters lase without phase-locking~\cite{Mahler20_3}. This is a qualitative but significant result: for low pump, although fewer lasers are for low pump compared to high pump, they are more consistently phase-locked. 

\section{Experimental and simulation Results}

\label{Experimental Results}

For full and quantitative characterization of the near-field behavior (percolation, cluster distribution, and lasing site correlations) and its far-field coherence, the site probability $p$ was varied from $0.3$ to $1$ with steps of $0.05$. 
For each $p$, near-field and far-field lasing intensity distribution were recorded for $50$ different random mask realizations. The percolation probability was calculated by measuring the percolating state of the array for each realization ($1$ if the edges of the laser array are connected by a single cluster and $0$ otherwise ~\citep{Stauffer94,StaufferAharony2018}) and then taking the ensemble average over the $50$ realizations. The number of clusters $N_{c}$ and the site correlation length $\xi$, i.e. the typical size or extent of connected clusters, were also determined from the measured near-field lasing intensity distributions.

\begin{figure}
\centering
\includegraphics[width=0.95\linewidth]{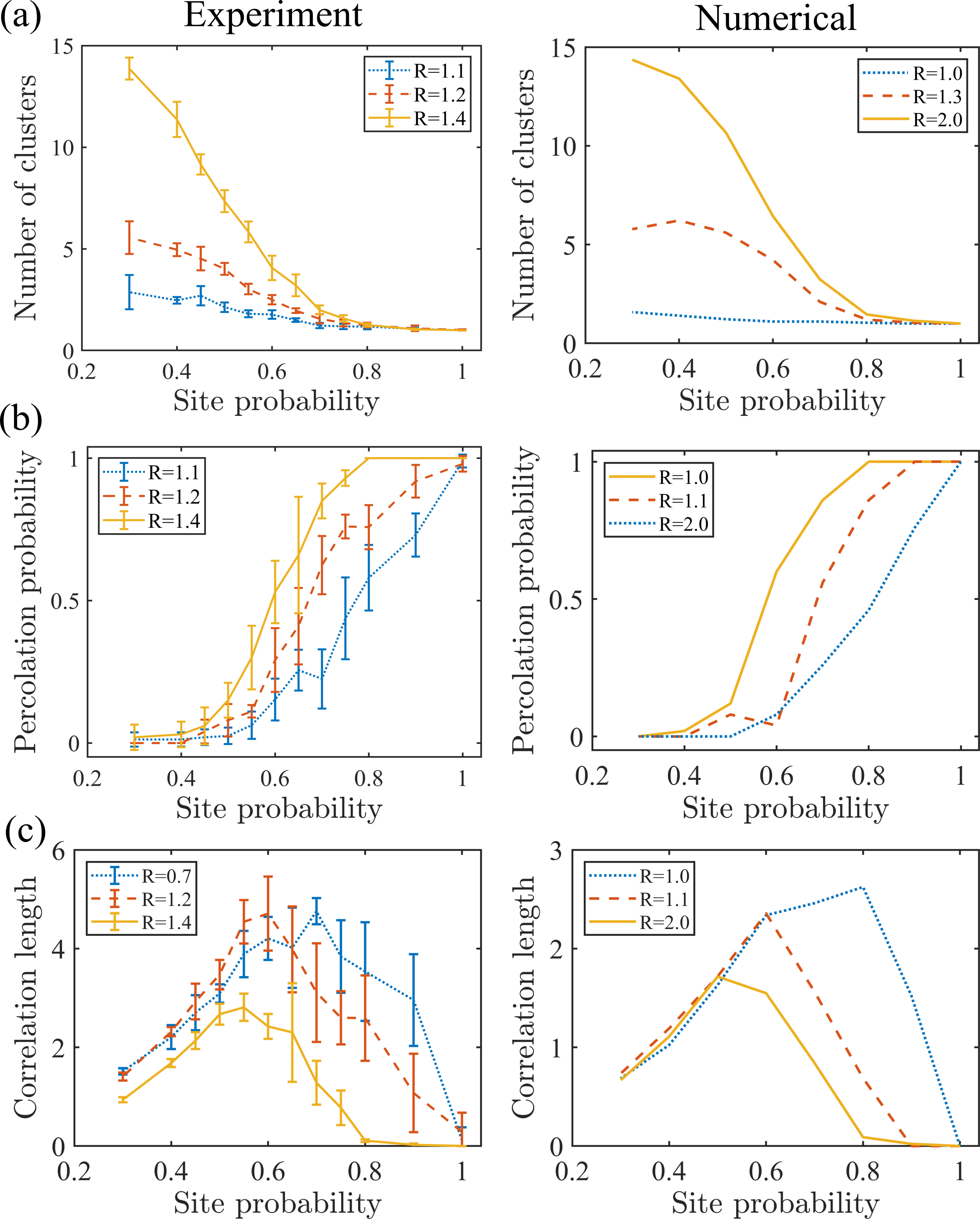}
\caption{Experimental and numerical percolation near-field results with coupled lasers as a function of the site probability $p$ for different pump ratios $R$. (a) Number of clusters,          (b) percolation probability and (c) site correlation length. All values are averaged over $50$ random mask realization. Error-bars are the standard deviation over the $50$ realizations. As seen, the number of clusters decreases with $R$ while the percolation transition threshold and the site correlation length increase, all due to the failure of smaller clusters to lase at small pumps.}
\label{fig:Fig_2_PercolationRes}
\end{figure}

Results of these near-field measurement are presented in Fig.~\ref{fig:Fig_2_PercolationRes} (left) for different pump ratio $R=P/P_{th}$ values (the ratio between the applied pump power $P$ and the pump power at lasing threshold $P_{th}$). 
Figure ~\ref{fig:Fig_2_PercolationRes} (a) left column shows  the measured number of clusters as a function of $p$ for 3 different pump strengths. For high pump ($R=1.4$), the number of clusters is maximal at $N_{c}=15$ for $p=0.3$ and then decreases as $p$ increases. All the lasers supported by the reflecting parts in the SLM's mask are found to lase. Thus, the number of clusters at high pump trivially matches the number of clusters generated in the SLM's mask (as in Fig.~\ref{fig:Fig_1_Exp_sketch}b, middle and Fig. S2 in Supplementary Material). 

At lower pump ($R=1.2$), the number of clusters, is reduced (e.g. $N_{c}=5$ at $p=0.3$), indicating that the smaller clusters generated in the SLM's mask cannot support lasing. At even lower pump ($R=1.1$), the number of clusters is  close to $N_{c}=1$ for all site probabilities $p$, indicating that near lasing threshold, only the largest clusters survive. For more details on the ratio of lasing sites as a function of the site probability $p$ see Fig. S2 in Supplementary Material.

The failure of the smaller clusters to lase at lower pump values has a dramatic affect on the percolation transition, as seen in Fig.~\ref{fig:Fig_2_PercolationRes} (b). While a percolation transition is clearly observed for all pump values, it occurs at higher probabilities $p$ as the pump value is reduced. This indicates that, as lasing threshold is approached, more lasers that belong to small isolated clusters experience higher losses and fail to lase. This, also increases the site correlation length (corresponding to the average cluster size) at lower pump ratios, as clearly seen in Fig.~\ref{fig:Fig_2_PercolationRes} (c). Note that the size of our laser array ($N = 100$) introduces finite-size effects: (1) the percolation transition is relatively smooth (gradual increase instead of a sharp transition) and (2) the apparent threshold is slightly shifted compared to the infinite-system \citep{Saberi15,Ziff2010}. Nevertheless, since all measurements are performed with the same array size, the observations regarding the nonlinearities shift remain valid.

We also performed numerical simulations of the laser dynamical equations to compare with our experiments. Simulations were carried out using an iterative Gerchberg–Saxton–Fox–Li algorithm \citep{Yang94,Zalevsky96,Huang21}. The simulation begins by initializing a laser field $E(x, y)$ over the transverse spatial coordinates $(x, y)$ with uniform amplitude and random phase. Each optical element of the DCL is then applied sequentially to this field as multiplicative or Fourier operators. Specifically, the spatial light modulator (SLM) transmission mask $T(x, y)$ is first applied, followed by the saturable gain function $G(x, y)$ defined as $G(x,y) = \frac{G_0}{1 + \|E_n(x,y)\|^2 / I_{\mathrm{sat}}}$, see Supplementary Material Section III. The field is then propagated through a Fourier-transforming lens, after which a coupling-lens function $CL(x, y)$ is imposed in the far-field. The resulting field is subsequently inverse-Fourier-transformed back to the near-field. This sequence constitutes a single cavity round trip, and the process is iteratively repeated. More details about the algorithm can be found in Ref.\cite{Naresh22} and in the Supplementary Material (where MATLAB code is also provided).

The numerical simulations, shown in the right columns of Fig.~\ref{fig:Fig_2_PercolationRes}, are in quantitative agreement with the experimental observations and provide a verification of our results. Both the reduction of the number of lasing clusters at lower pump values, as well as the systematic shift of the percolation transition toward higher site probabilities $p$ when approaching the lasing threshold are clearly observed in the numerical simulations. We note that in the simulations we use slightly different values of the pump ratio $R$ compared to the experiment. This is because, in the intermediate regime, the percolation transition could not be determined with a single fixed value of $R$.

Next, we quantified the quality of phase-locking of the coupled lasers (i.e. their coherence) by analyzing the measured far-field intensity distributions ~\citep{Mahler19_2,Mahler20_3}. Specifically, we calculated the number of lasers phase-locked $M_{locked}$ using Eq.\ref{eq:M_locked}, and the inverse participation ratio (IPR) , defined as~\citep{Mahler20_3} $\frac{\left(\sum_i I^{ff}_i\right)^2}{\sum_i (I^{ff}_i)^2}$.
The far-field IPR is a standard measure of localization and is commonly used to quantify phase synchronization among oscillators~\citep{Witthaut14}. 

\begin{figure}
\includegraphics[width=0.95\linewidth]{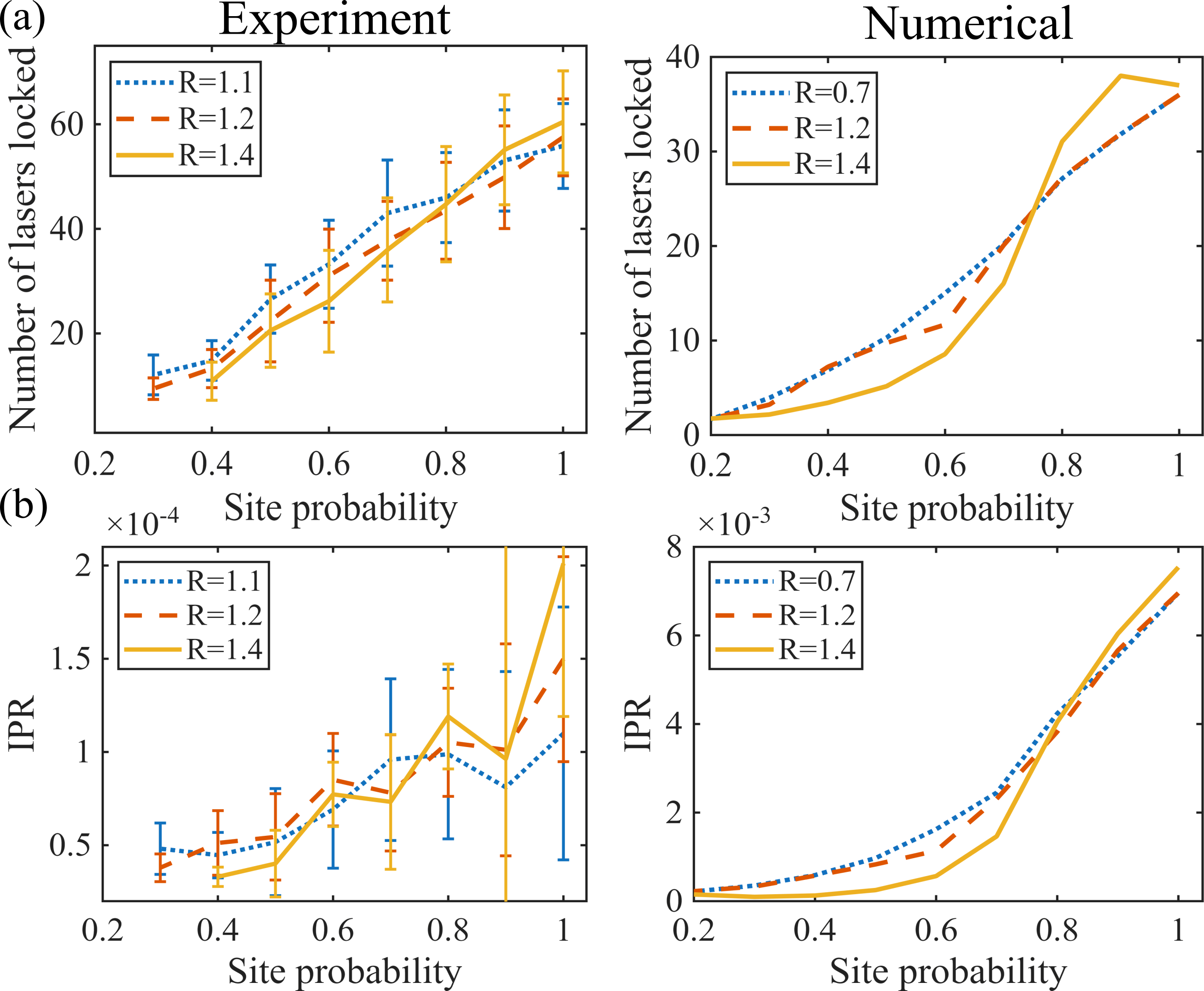}
\caption{Experimental and numerical far-field  results  of percolating coupled laser array characterizing its coherence. (a) Number of phase-locked lasers $M_{locked}$ and (b) inverse participation ratio (IPR) as a function of the site probability $p$ for different pump ratio $R$. All values are averaged over 50 random mask realization and error-bars are the standard deviation over the $50$ realizations. As seen, at low $p$ both $M_{locked}$ and the IPR indicate slightly higher coherence for the smaller pump ratios. As $p$ increases, coherence improves more at higher pump ratio and overtakes the low pump  coherence.}
\label{fig:Fig_3_M_locked}
\end{figure}

The experimental coherence results are presented in Fig.~\ref{fig:Fig_3_M_locked} (left). At low site probability $p$, both $M_{locked}$ and the IPR indicate slightly higher coherence for the smaller pump ratios $R$. As the site probability $p$ increases, coherence improves more at higher pump ratios and is higher than the low pump  ratio coherence—around $p\approx0.75$. Although marginal, this trend is also clearly observed in the numerical simulations of Fig.~\ref{fig:Fig_3_M_locked} (right). 


Note that the reported phase locking metrics are extracted from far-field diffraction peaks~\citep{pando2024synchronization,Naresh22,Mahler20}, whose sharpness and contrast reflect the degree of phase locking. The imaging system employed in this study resolved these far-field peaks, ensuring that the measurement is not limited by spatial resolution. The camera exposure time integrated over the full laser pulse duration, capturing all temporal and spatial modes. The large spread in the experimental data (error-bars in Fig. 3) is primarily attributed to the intrinsic sensitivity of phase locking to noise and system fluctuations. Specifically, phase locking in laser arrays is highly sensitive to perturbations such as cavity length fluctuations (as small as $\lambda/50$), multimode dynamics~\citep{Mahler21,Mahler22}, and pump fluctuations~\citep{pando2024synchronization,Naresh22,Mahler20}, which directly affect the stability and visibility of phase-locked states.

\section{Theoretical Toy Model}
\begin{figure}
\includegraphics[width=0.95\linewidth]{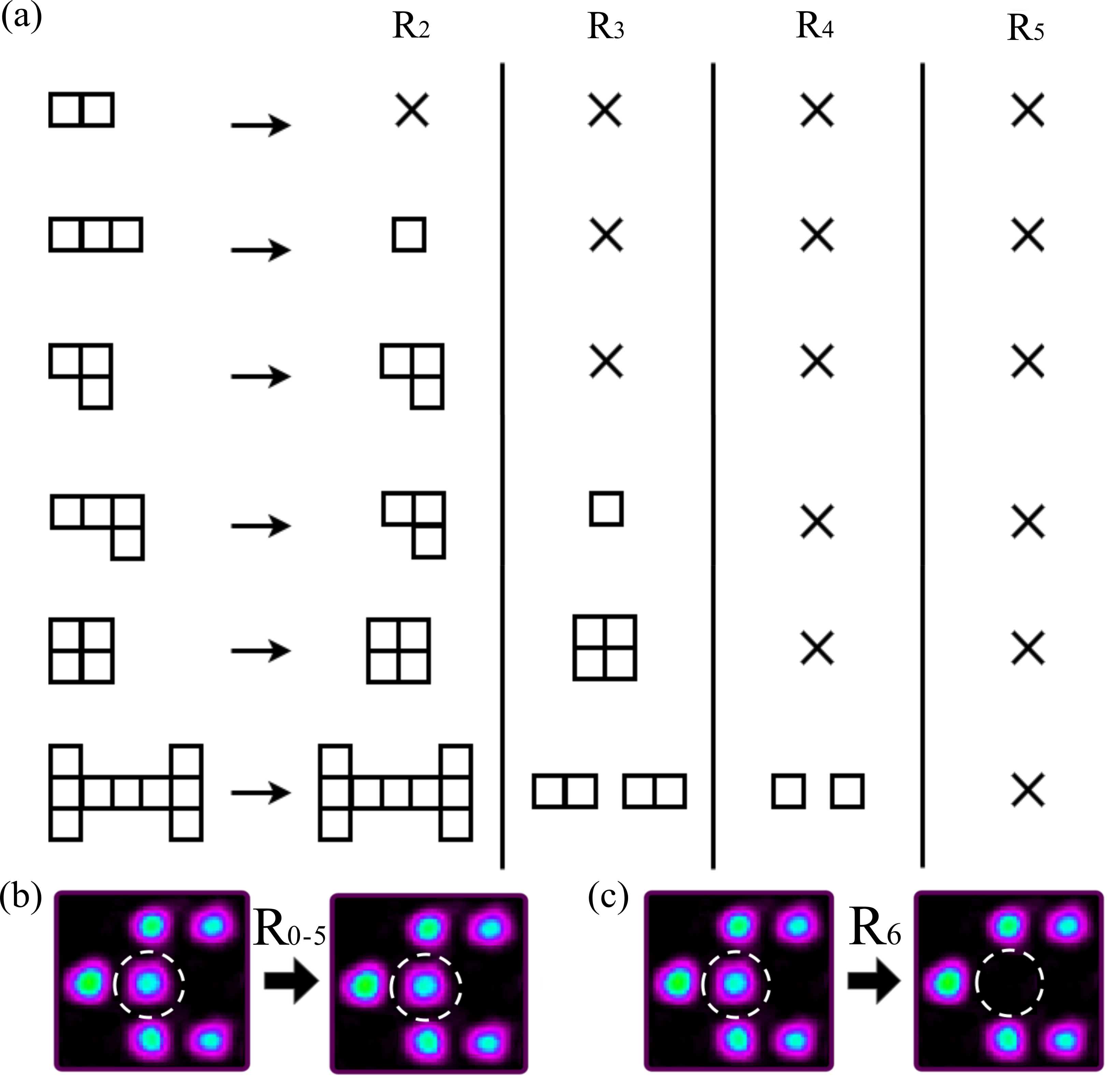}
\caption{(a) Schematic representation of rules $R_{2}$ to $R_{5}$ applied to lattice animals (polyominoes) of different size. (b) and (c) Representation example of rules $R_{1}$ to $R_{5}$  and $R_{6}$ applied to a laser site in a cluster.}
\label{fig:rule_explanation}
\end{figure}

To interpret the experimental and simulation results of Fig.~\ref{fig:Fig_2_PercolationRes}, we resorted to a simple theoretical percolation survival-game toy model (Fig.~\ref{fig:rule_explanation}) that involves the statistical behavior of site occupation and allows us to contextualize the observed transition and its dependence on the pump level. It serves as a model providing qualitative intuitions to the underlying laser dynamics in percolation. A derivation that connects the survival-game toy model framework to the laser dynamics through a sequence of approximations is presented in the Supplementary Material (Part III: Hierarchy of Models: From Experiment to Theory). In the toy model, we include rules denoted as $R_N$. These rules define a survival game for occupied sites, and are intended to mimic the lasing nonlinearity effects observed with the coupled lasers system. 

A rule $R_{N}$ (ranging from $R_{0}$ to $R_{8}$) requires that each site in the lattice must have at least $N$ occupied neighbors among the eight neighbors of a square lattice to survive; otherwise, the site is switched off. For example, rule $R_{2}$ requires that an occupied site  have at least two occupied neighboring sites to remain occupied. If an occupied site has zero or one occupied neighbor, it becomes unoccupied in the next step, as shown in Fig.~\ref{fig:rule_explanation}. This update is performed in parallel for all sites. As $N$ increases, the rule more strongly alters the lattice by removing small clusters. 
To mitigate boundary artifacts near the edges of the lattice, the lattice is extended by adding additional edge lines during the rules application and removing them before the final update.

Rule $R_{0}$ ($N=0$) corresponds to a site that does not require neighbors to survive; consequently, all sites survive. This case is equivalent to conventional classical percolation. At the other extreme, rule $R_{8}$ ($N=8$) corresponds to a site that requires all eight neighboring sites (which is the maximum possible number of neighbors on a square lattice) to survive.

One can combine these effective rules with classical percolation theory, using the standard lattice animals expansion:
\begin{equation}
n_s = \sum_t n_{st} = \sum_t g_{st} p^s (1 - p)^t,
\label{eq:lattice_animal}
\end{equation}
which gives the cluster-size distribution: the average number of clusters of size $s$ (i.e., clusters of $s$ lasers in our case) per lattice site at occupation probability $p$. This expansion accounts for all clusters of a given size $s$ with different configurations, characterized by their perimeter $t$. Here, $g_{st}$ is the number of distinct lattice animals (cluster configurations) with size $s$ and perimeter $t$.

The rules act as nonlinear “survival game” operators acting on $n_{st}(p)$, mimicking the nonlinearities induced by the experimental pump-ratio parameters $R$. Their are implemented  as (see polyominoes illustrated in Fig.~\ref{fig:rule_explanation}) as
\begin{equation}
\tilde{n}_j(p) = R^N_{ji} n_i(p),
\label{transformation}
\end{equation}
where $R^N_{ji}$ is the matrix representation of rule $R_N$ and $\tilde{n}_j(p)$ denotes the updated state. We emphasize that this is not an exact description of the laser system, but rather a coarse approximation. Indeed, the model complexity was reduced by using intensity representation $|E|^2$ (instead of complex field) and by restricting the parameter space to a binary mask configuration $m_i$ and the pump strength $R$. This leads to a coarse-grained, site-resolved description from which observables such as cluster statistics and correlation length can be extracted. The continuous dependence of the intensity on $R$ is then mapped onto a discrete ``survival rule,'' where each site is assigned a binary occupation variable $n_i$ depending on whether its intensity exceeds a threshold. While approximate, this construction provides a link between the underlying laser dynamics and the emergent percolation behavior, capturing the key qualitative features. A more detailed derivation connecting this framework to the laser system is provided in the Supplementary Material Section V. Hierarchy of Models: From Experiment to Theory.


\begin{figure}
\includegraphics[width=0.95\linewidth]{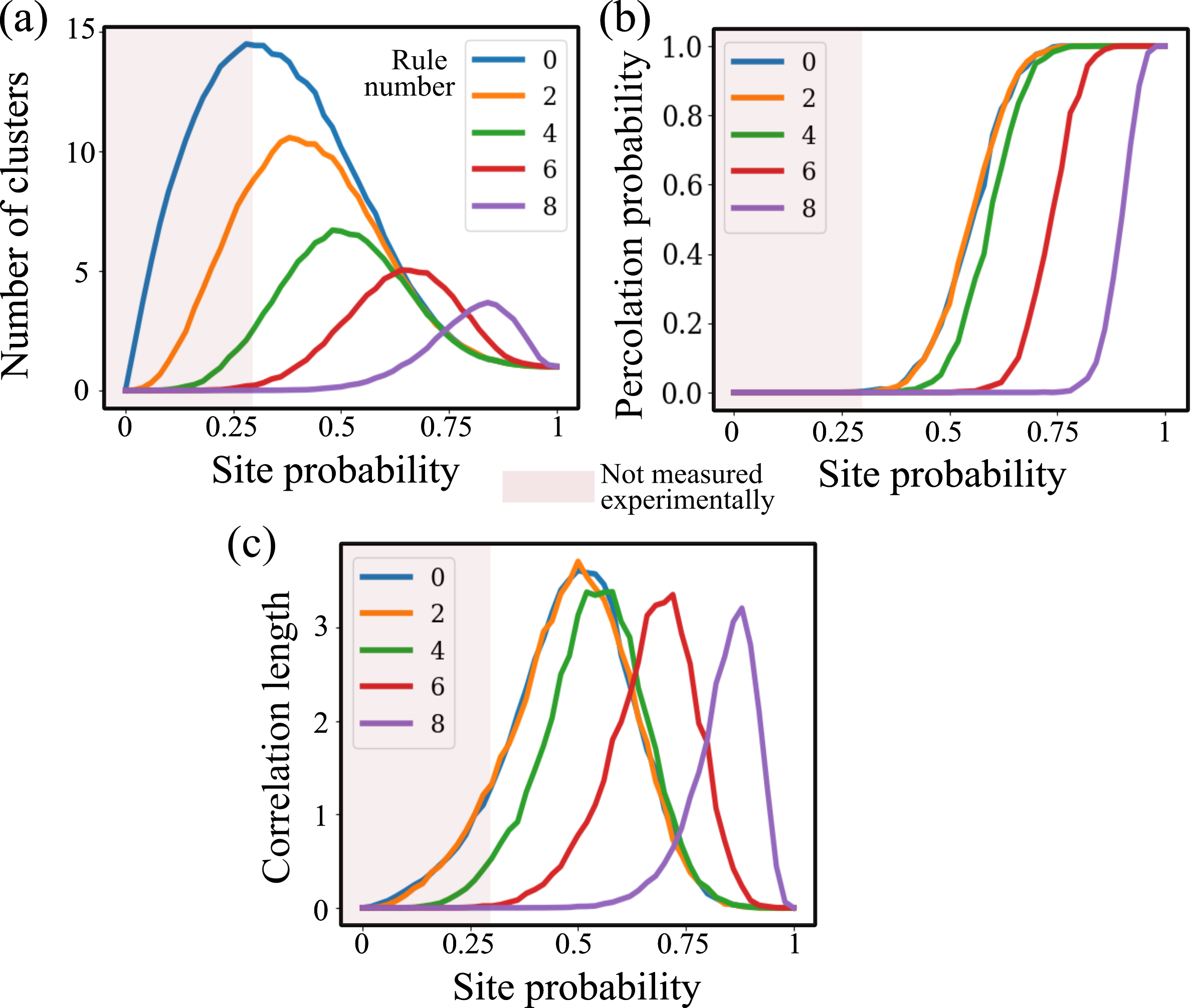}
\caption{Results of a theoretical percolation toy-model that maps the pump strength of the sites onto a discrete neighbor-dependent survival rule. (a) Number of clusters, (b) percolation probability, and (c) correlation length as functions of the site occupation probability $p$ for the different rules $R_{N}$ applied to classical percolation of $M=100$ sites square array. The red shaded region indicates the range of $p$ not explored experimentally in Fig. 2.}
\label{fig:Fig_5}
\end{figure}

The calculated average results over $4000$ random realizations of our toy model are presented in Fig.~\ref{fig:Fig_5} for a square lattice of $100$ sites. As evident, increasing the rule number applied to the percolation lattice (a) decreases the number of clusters for small $p$, (b) delays the percolation transition to higher values of $p$, and (c) shifts the maximal peak of the site correlation length to larger values of $p$. All these are in qualitative agreement with our experimental and simulation results of the percolating coupled laser array of Fig.~\ref{fig:Fig_2_PercolationRes}, indicating that increasing the survival rule number $N$ can qualitatively mimic the increase in nonlinearity as the pump is reduced towards lasing threshold. 


\section{Conclusions}
\label{Conclusions}
We developed a tunable and programmable photonic platform for studying nonlinear percolation using a two-dimensional square array of $100$ coupled lasers. The emergence of a percolating clusters coincides with global phase locking of the laser array and follows classical percolation behavior at high pump powers where the role of nonlinearities does not play a role. Surprisingly, at low pump power, nonlinearities (due to more selective mode competition) shift the percolation threshold to higher values of site probability $p$, and modify the cluster size distribution. It also shifts the correlation length by selectively suppressing small clusters. These survival nonlinearities act in the opposite sense to conventional laser nonlinearities (e.g., gain saturation or Kerr-type effects): where at high pump levels, the system approaches classical percolation behavior, and at low pump levels, deviations become more pronounced (see Supplementary Material Section II). Finally, this nonlinear regime also improved the quality of phase locking without extending its range. 

To interpret the experimental and numerical results, we introduced a neighbor-dependent connectivity-rule survival model that can account for nonlinear effects. Simulations based on this toy model reproduced the main qualitative features of the experimental results. Specifically, we demonstrated that increasing the survival rule number $N$ mimicked the nonlinear regime (Supplementary Material). While the rules primarily shift the percolation threshold, they do not alter the asymptotic behavior of the clusters distribution, see Supplementary Material Section V. Critical Indices. Future studies could characterize and quantify the nature of the transition with the extraction of critical exponents and finite-size effects accounting.

A key feature of our photonic platform is the ability to simultaneously access intensity-based percolation and phase-locking transitions. Unlike other platforms that address either geometric percolation or phase synchronization separately, ours deals with both simultaneously via near-field and far-field  measurements. This enables a deeper exploration of how connectivity and nonlinearities jointly govern the emergence, quality, and robustness of coherence near criticality. A second key feature is the "survival game" revealed by controlling the laser pumping, where the nonlinear gain selectively suppresses weakly connected clusters, modifying both the percolation threshold and global coherence. 

These features extend recent progress in analog photonic computing and alternative photonic simulators~\cite{Mahler25,stroev2023analog, stroev2021discrete}. Our reconfigurable optical percolation platform could lead to a wide range of practical simulations and applications. It could serve as a metamaterials simulator, modeling tunable disorder and connectivity in artificial optical media. The site-dependent activation and nonlinear control could be exploited for complex network phenomena~\citep{Souza2019,Lee2018}, including forest fire propagation~\citep{WVonNiessen_1986,Perestrelo2022,vandenBerg2021}, where increased “pump” mimics environmental stress or dryness; disease spreading dynamics, where site occupation and phase coherence correspond to infection probability and the synchronization of outbreaks; a testbed for resilience and collapse in active networks, such as neural activity patterns, power grids, or ecosystems under stress. The unique combination of tunable geometry, nonlinear suppression, and coherence tracking may enable exploration of complex percolation-like transitions in a variety of real-world systems.

Our approach emphasizes that percolation is fundamentally a physical process defined by connectivity, correlations, and collective transitions, which can be directly mapped onto a physical optical system. In this sense, the laser array does not compute percolation in a conventional sense, but rather realizes it physically, where connectivity, noise, and nonlinearities are inherently present and co-evolve with the system dynamics. Such physical realization is particularly relevant in the context of optical and photonic computing, where information processing is increasingly explored through distributed, analog, and nonlinear media rather than purely digital architectures. In this framework, the advantage of the optical platform is not computational flexibility, but the ability to naturally embed and observe coherence, fluctuations, and nonlinear suppression mechanisms that are intrinsic to the physical system itself.

\begin{acknowledgments}
The authors thank the Israel Science Foundation (funding no. 501100003977) for their support. Nikita Stroev acknowledges the support from the Weizmann–UK Make Connection grant (grant no. 142568).
\end{acknowledgments}

\section*{Data Availability Statement}
The data that support the findings of this study are available from the corresponding author upon reasonable request.

\section*{References}
\nocite{*}
\bibliography{bibliography}
\end{document}